\documentclass[12pt,a4paper]{article}
\usepackage{epsfig}
\usepackage{graphics}

\newcommand{\etal} {et al.}

\usepackage{amssymb}
\usepackage{epsfig}

\newcommand{\mm}[1]{{\mbox{\hspace{#1mm}}}}
\newcommand{\beq}[1]{\begin{equation}{\label{#1}}}
\newcommand{\eeq}[0]{\end{equation}}
\newcommand{\eq}[1]{Eq.~(\ref{#1})}

\newcommand{\fig}[1]{Fig.~\ref{#1}}
\newcommand{\Fig}[1]{Figure~\ref{#1}}
\newcommand{\vev}[1]{\langle #1 \rangle}

\begin{document}
\title{A statistical method for luminosity monitoring in high energy
  collider experiments}

\author{Jo\~ao Bastos$^1$, Jo\~ao Carvalho$^1$, Michael Schmelling$^2$ \\
{\small {\sl $^1$LIP-Coimbra, Univ. de Coimbra, P-3004-516 Coimbra, Portugal}} \\
{\small {\sl $^2$Max-Planck-Institut f\"ur Kernphysik, D-69029 Heidelberg, Germany }}
}
\date{ }
\maketitle

\begin{abstract}
A statistical method which uses a combination of two subdetectors to monitor
the luminosity in high energy interactions is presented. To illustrate
its performance, this method was applied to random triggered minimum bias 
data collected in the commissioning period of the \mbox{HERA-B} experiment 
in spring 2000. It is found that luminosity estimates with an intrinsic 
systematic error of 3\% can be obtained.
\end{abstract}

%%%%%%%%%%%%%%%%%%%%%%%%%%%%%%%%%%%%%%%%%%%%%%%%%%%%%%%%%%%%%%%%%%%%%%%%%%
%%                            INTRODUCTION                              %%
%%%%%%%%%%%%%%%%%%%%%%%%%%%%%%%%%%%%%%%%%%%%%%%%%%%%%%%%%%%%%%%%%%%%%%%%%%

\section{Introduction}
\label{introduction}

The precise determination of the luminosity of experimental data is required
for absolute cross section measurements. Luminosity $l$\/ is defined as the 
proportionality factor between interaction rate and cross section for the 
process under consideration. The integrated luminosity $L$\/ relates cross 
section $\sigma$\/ and interaction count $N$\/ for a time interval $T$, i.e.
\beq{lumidef}
   l(t) = \frac{1}{\sigma} \frac{dN}{dt}
   \mm{10}\mbox{and thus}\mm{10}
   L=\int_0^T l(t)dt = \frac{N}{\sigma} \;.
\eeq

Given the cross section $\sigma$\/ for a particular process, such as the 
inelastic cross section in high energy hadronic interactions, the
determination of the integrated luminosity for a given data set is 
equivalent to determining the number of interactions of that process. 
For the following we will focus on collider experiments, where a bunched
beam produces events with a well defined time structure, and where the 
number of interactions per bunch crossing will fluctuate statistically. 

As a first approach, determining the number of interactions could be 
accomplished by simply counting the number of reconstructed primary 
vertices in the data. To achieve that, the vertex reconstruction 
efficiencies must be known. Additionally, two neighboring vertices may 
be merged by the reconstruction package while others may be split. 
Therefore, the probabilities for these processes must also be calculated, 
which is often difficult and introduces poorly known systematic errors. 
An alternative technique consists in extracting the total number of 
interactions from inclusive quantities in the reconstructed data 
which are proportional to the number of primary collisions in an 
event, such as the number of hits or the total energy deposition. 
Obviously, the main difficulty associated with this approach is the 
need for an absolute calibration, that is, the average signal for 
a single interaction must be known.

In a different approach (the so called``statistical method''), a poissonian 
distribution for the number of interactions per bunch crossing is assumed and 
the average number of interactions is extracted from the number of empty
events in the data sample. The advantage of this method is that nothing about 
the average signal for a single interaction has to be known or assumed. 
However, the acceptances for tagging non-empty events must be estimated and 
the occurrence of noise events, which may be tagged as non-empty, must be 
taken into account.

In this paper a method for determining the integrated luminosity by counting 
the fraction of empty events simultaneously in two subdetectors is proposed. 
With this procedure the detector acceptances for a single interaction and the 
fraction of noise events can in principle be obtained from data, relaxing 
the dependence on Monte Carlo simulations to derive these quantities and the 
introduction of systematic errors which are difficult to estimate.

This paper is organized as follows. In the next section, an expression for 
the probability to observe an empty event is derived, assuming that the
distribution of the number of interactions per bunch crossing follows Poisson
statistics but allowing also for non-negligible rate fluctuations. Based on 
this, in Section 3, the Two-System Statistical Method (TSSM) is introduced. 
It is shown how counting the fraction of empty events in either of two 
subdetectors and simultaneously in both allows to determine the acceptance 
of both subdetectors and the mean number of interactions in the data. In 
Section 4, this procedure is applied to minimum bias events collected in 
the commissioning period of the HERA-B experiment \cite{Hartouni} in 
spring 2000. The conclusions are presented in Section 5.

%%%%%%%%%%%%%%%%%%%%%%%%%%%%%%%%%%%%%%%%%%%%%%%%%%%%%%%%%%%%%%%%%%%%%%%%%%
%%                  COUNTING THE NUMBER OF EMPTY EVENTS                 %%
%%%%%%%%%%%%%%%%%%%%%%%%%%%%%%%%%%%%%%%%%%%%%%%%%%%%%%%%%%%%%%%%%%%%%%%%%%

\section{Counting Empty Events}
%------------------------------
A particle collider usually has circulating beams with many bunches 
contributing to the observed interaction rate. Since the individual bunch 
currents, in general, can differ by significant amounts, the following 
analysis is formulated for an ensemble of distinguishable bunches. This
entails a slight complication of the formalism, but, as will become clear
later, gains a lot of information which can be exploited in the analysis. 

For the start let us assume that the distribution of the number of
interactions per bunch crossing follows Poisson statistics. If the average 
number of interaction produced by bunch number $i$\/ is $\mu_i$, the 
probability that there are $n$\/ interactions in an event from this bunch 
crossing is
\beq{poisson}
\mathcal{P}(n,i)=\frac{\mu_i^n}{n!}e^{\displaystyle -\mu_i}.
\eeq
Now suppose that a certain subdetector is used to count the number of
empty events in the data set. An event is {\sl tagged} as being empty 
if a quantity associated with this subdetector (e.g. hits, tracks, energy 
deposition, etc.) is below a specified threshold value. This value represents 
a compromise between a large efficiency for tagging non-empty interactions 
and an effective exclusion of noisy ``events'' in which no interaction has 
occurred. The probability to observe an empty event in this system is
\beq{pnullgen}
P(0,i) = (1-q) \sum_{n=0}^{\infty}(1-a^{(n)}) \mathcal{P}(n,i),
\eeq
where $a^{(n)}$\/ is the acceptance, or efficiency, to tag an event with 
$n$\/ interactions as non-empty and $q$\/ is the probability to observe 
an event due to noise in the subdetector or to background 
(i.e. beam gas interactions). If the probability to pass the 
tagging threshold is independent of the number of primary interactions,
which to a good approximation is valid if the threshold is set such 
that a single interaction has a large probability to exceed it, 
then $a^{(n)}$\/ can be approximated by
\beq{acck}
a^{(n)} = 1 - (1-a)^n
\eeq
where $a \equiv a^{(1)}$ is the efficiency to tag a single interaction
as non-empty. Substituting \eq{poisson} and (\ref{pnullgen}) into 
\eq{acck} one gets
\beq{pnull}
P(0,i) = (1-q) e^{\displaystyle -a\mu_i}.
\eeq

However, some bunches may suffer from rate instabilities so that 
\eq{poisson} does no longer describe the interaction multiplicities
correctly. In this case, the average number of interactions $\mu_i$\/ 
is no longer constant but it fluctuates by a random amount $\nu_i$\/ 
around its central value, $\mu_i \rightarrow \mu_i + \nu_i$. With
$g(\nu_i)$\/ the probability density function of those fluctuations,
the probability to observe an empty event becomes
\beq{ratefluctgen}
   P(0,i) =  
       (1-q)\int_{-\mu_i}^\infty e^{\displaystyle -a(\mu_i+\nu_i)} 
        g(\nu_i)d\nu_i
\eeq
Assuming further that the fluctuations around $\mu_i$\/ are Gaussian 
distributed, with zero average $\vev{\nu}_i=0$\/
and standard deviation $\sigma_i\ll \mu_i$, \eq{ratefluctgen} can be 
integrated analytically to yield
\beq{ratefluct}
P(0,i) = (1-q)\exp\left(-a\mu_i+\frac{1}{2}a^2\sigma_i^2\right) \;.
\eeq
One sees that rate fluctuations enter as second order effects, i.e. as
long as they are small the assumption of poissonian distribution for
interaction multiplicities is a good approximation. Large rate 
fluctuations, however, have a sizeable impact and have to be taken
into account in the analysis.

%%%%%%%%%%%%%%%%%%%%%%%%%%%%%%%%%%%%%%%%%%%%%%%%%%%%%%%%%%%%%%%%%%%%%%%%%%
%%                  THE TWO-SYSTEM STATISTICAL METHOD                   %%
%%%%%%%%%%%%%%%%%%%%%%%%%%%%%%%%%%%%%%%%%%%%%%%%%%%%%%%%%%%%%%%%%%%%%%%%%%

\section{The Two-System Statistical Method}
%------------------------------------------
Now let us consider {\sl two} subdetectors or combinations of subdetectors,
which will be denoted by ``system 1'' and ``system 2''. According to 
\eq{ratefluct} the probabilities $p_k$, $k=\{1,2\}$\/ to observe an 
empty event in either of the two systems are
\beq{pnullk}  
     p_{k\,i} \equiv P(0,i)_k 
 = (1-q_k)\exp\left(-a_k\mu_i+\frac{1}{2}a_k^2\sigma^2_i\right)
\eeq
where $q_k$\/ is the probability to record an event due to background 
or noise in system $k$\/ and $a_k$\/ is the efficiency to tag 
single interactions in this system. If the two systems are independent 
the probability $p_0$\/ to observe an empty event simultaneously in both 
subdetectors is given by an analogous expression
\beq{pnullc}
  p_{0\,i} = (1-q_0)\exp\left(-a_0\mu_i+\frac{1}{2}a_0^2\sigma^2_i\right) \;,
\eeq
where $q_0 = q_1+q_2-q_1q_2$\/ and $a_0 = a_1+a_2-a_1a_2$.

In order to get a handle on rate fluctuations, we now combine the statistical 
approach with a measurement based on an inclusive quantity, which is
insensitive to deviations from a poissonian for the interaction multiplicities.
This is achieved by expressing $\mu_i$\/ in terms of a bunch dependent 
inclusive quantity $\vev{n}_i$\/ which is proportional to the number of 
interactions per bunch crossing,
\beq{taudef}
  \vev{n}_i = \tau \, \mu_i \;.
\eeq
The parameter $\tau$\/ is the mean value of the inclusive quantity per 
interaction within the detector acceptance. Substituting \eq{taudef} 
in \eq{pnullk} and (\ref{pnullc}) we have
\beq{pmaster}
   p_{k\,i}  
 =  (1-q_k)\exp\left(-a_k\frac{\langle n \rangle_i}{\tau}+\frac{1}{2}a_k^2\sigma^2_i\right),
   \mm{5}\mbox{with}\mm{5} 
   k=\{0,1,2\}\;.
\eeq

If there are no rate fluctuations, $\sigma_i \simeq 0, \forall i$, the global 
(bunch independent) parameters, $q_k$, $a_k$\/ and $\tau$, can be obtained
from \eq{pmaster} by fitting the values of $p_{k\,i}$\/ as a function of 
the observable $\vev{n}_i$. Once $\tau$\/ is known, the average number of 
interactions $\mu_i$\/ for every bunch is calculated according to
\eq{taudef}. From $\mu_i$\/ and the number of recorded events in each
of the bunches, the total number of interactions in the data sample and thus
the integrated luminosity can be calculated.
 
In case that rate fluctuations are present for some bunches, those bunches 
have to be identified and removed from the global fit. This can be achieved 
by considering the following relation between the probabilities $p_{ki}$,
\beq{discr}
\ln{\frac{p_{0i}}{p_{1i}p_{2i}}}=a_1a_2\left[\frac{\langle n \rangle_i}{\tau}
+\sigma_i^2 \left(1 - a_0 - \frac{1}{2}a_1a_2 \right)\right] \;.
\eeq
In case of negligible rate fluctuations we have a simple linear relation between
$\ln (p_{0i}/p_{1i}p_{2i})$\/ and $\vev{n}_i$. Bunches with significant rate
fluctuations would deviate from that relation and can be excluded from the 
global fit. Note that \eq{discr} also has the potential to detect situations
where all bunches are subject to rate fluctuations. In this case one has 
no outlier bunches, but a straight line fit to $\ln (p_{0i}/p_{1i}p_{1i})$\/ 
versus $\vev{n}_i$\/ would not pass through the origin, unless
$\sigma_i^2$\/ and $\mu_i$\/ are proportional.

%%%%%%%%%%%%%%%%%%%%%%%%%%%%%%%%%%%%%%%%%%%%%%%%%%%%%%%%%%%%%%%%%%%%%%%%%%
%%                    AN APPLICATION TO HERA-B DATA                     %%
%%%%%%%%%%%%%%%%%%%%%%%%%%%%%%%%%%%%%%%%%%%%%%%%%%%%%%%%%%%%%%%%%%%%%%%%%%

\section{Application to HERA-B Minimum Bias Data}
%------------------------------------------------
To illustrate its properties, the proposed method was applied to
minimum bias events, collected with a simple random trigger during 
the HERA-B commissioning period in spring 2000. Applying the TSSM to 
real data shows how the considerations that went into its design cope 
with problems arising under realistic conditions. 
With respect to HERA-B, please note that the random-trigger based method 
described in this paper should not be confused with other ones employed 
by the HERA-B collaboration for luminosity measurements, such as for 
example the method \cite{marco} applied to the interaction-triggered data 
recorded in 2002/2003. 

HERA-B is a large acceptance fixed target experiment that studies the 
interactions of 920~GeV protons with wire targets placed in the beam halo 
of the HERA storage ring, at DESY. The HERA-B target \cite{Target} consists 
of two stations separated by 4~cm along the beam. Each station comprises 
four wires of different materials, with dimensions ranging from 0.5 to 1~mm 
along the beam and from 50 to 100~$\mu$m perpendicular to the beam. Each 
wire can be independently moved inside the beam halo in order to adjust 
the interaction rate. The reconstruction of primary and secondary vertices 
is performed by a silicon micro-strip Vertex Detector System \cite{VDS}. 
The main tracker is divided into the Inner Tracker \cite{ITR}, composed 
of micro-strip gas chambers with gas electron multipliers, and the Outer 
Tracker \cite{OTR} made of honeycomb drift cells. Particle identification 
is performed by a ring imaging Cherenkov detector \cite{RICH}, an 
electromagnetic calorimeter \cite{ECAL} and a muon detector \cite{MUON}.

The runs analysed in the following were taken with four different target 
materials: carbon, aluminum, titanium and tungsten. The nominal interaction rates 
ranged from 2 to 20 MHz. In the HERA proton ring there are 220 slots for 
bunches separated by 96 ns. Usually only 180 of these are filled with protons. 
These are organized in three groups of 60 bunches, separated by three gaps of \mbox{5+5+15}
empty slots. In turn, each group is composed of 6 subgroups of 10 contiguous bunches.
These subgroups are separated by a single empty slot.

\begin{figure}[htb]
\centering
\epsfig{file=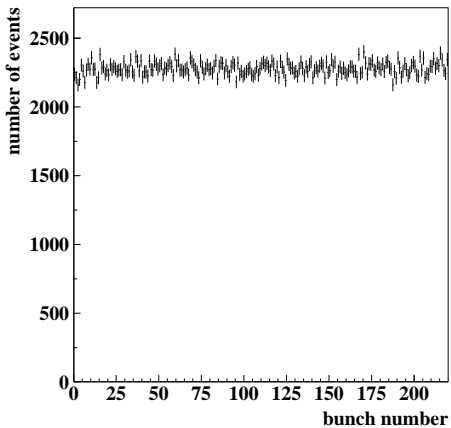, width=6cm}
\caption{Number of events recorded with a random trigger as a function of bunch number.
Bunches that are nominally empty are also sampled.}
\label{bxdist}
\end{figure}

In \fig{bxdist}  one can see the number of recorded events as a functions of the 
bunch number for a run of 500k events taken with a carbon wire target. 
As can be seen, the data acquisition system samples all bunches very 
uniformly (even the ones which are nominally empty). 

At HERA-B inelastic interactions dominate the total visible cross section 
and therefore they are the natural reference process for luminosity 
determination. Elastic scattering events are normally outside the detectors
acceptances and contribute marginally to the rate. The inelastic cross 
section of $pA$\/ collisions was measured by several fixed target
experiments, for a large number of target materials and beam energies
\cite{murthy,bellettini,carroll,fumuro,denisov,roberts}. It is found to 
be approximately independent of the incident particle energy and a power 
law dependence on the target atomic weight $A$\/ is well fitted by the 
experimental data \cite{Carvalho}. The inelastic cross section comprises 
a non-diffractive and a diffractive component. Since for the latter
both experimental acceptance and the contribution to the total cross
section are small, it is a good approximation to assume that only the 
non-diffractive component of the inelastic cross section contributes
to the luminosity determination. The resulting bias can be estimated 
by Monte Carlo simulations.

\subsection{Mean number of tracks per bunch crossing}
%----------------------------------------------------
In \eq{taudef} $\mu_i$\/ was expressed in terms of an inclusive quantity
which is proportional to the number of interactions per bunch crossing. 
In the following we choose this quantity to be the mean number of reconstructed 
tracks $\vev{n_t}$\/ which, to a good approximation, scales linearly with
the number of primary collisions, i.e. $\vev{n_t}= \tau \mu_i$, where
$\tau$ is the mean number of reconstructed tracks in one interaction.
The validity of this assumption was checked with a Monte Carlo simulation
based on the FRITIOF 7.02 generator \cite{FRITIOF} and the subsequent 
simulation of the HERA-B detector. In order to exclusively select tracks 
originating from primary interactions, and eliminate non-target related 
tracks from secondary decays such as $K_S\rightarrow\pi^+\pi^-$\/ and 
conversions $\gamma\rightarrow e^+e^-$, the following selection criteria 
were applied to all tracks in the event. Only tracks containing at least 
6 reconstructed hits in the vertex detector (VDS) are accepted. To avoid 
counting multiply reconstructed tracks (the so-called {\sl clones}),
tracks sharing a VDS segment with a previously accepted track were rejected.
Finally, an impact parameter below 1 ~mm at the primary vertex is required.

\begin{figure}[htb]
\centering
\epsfig{file=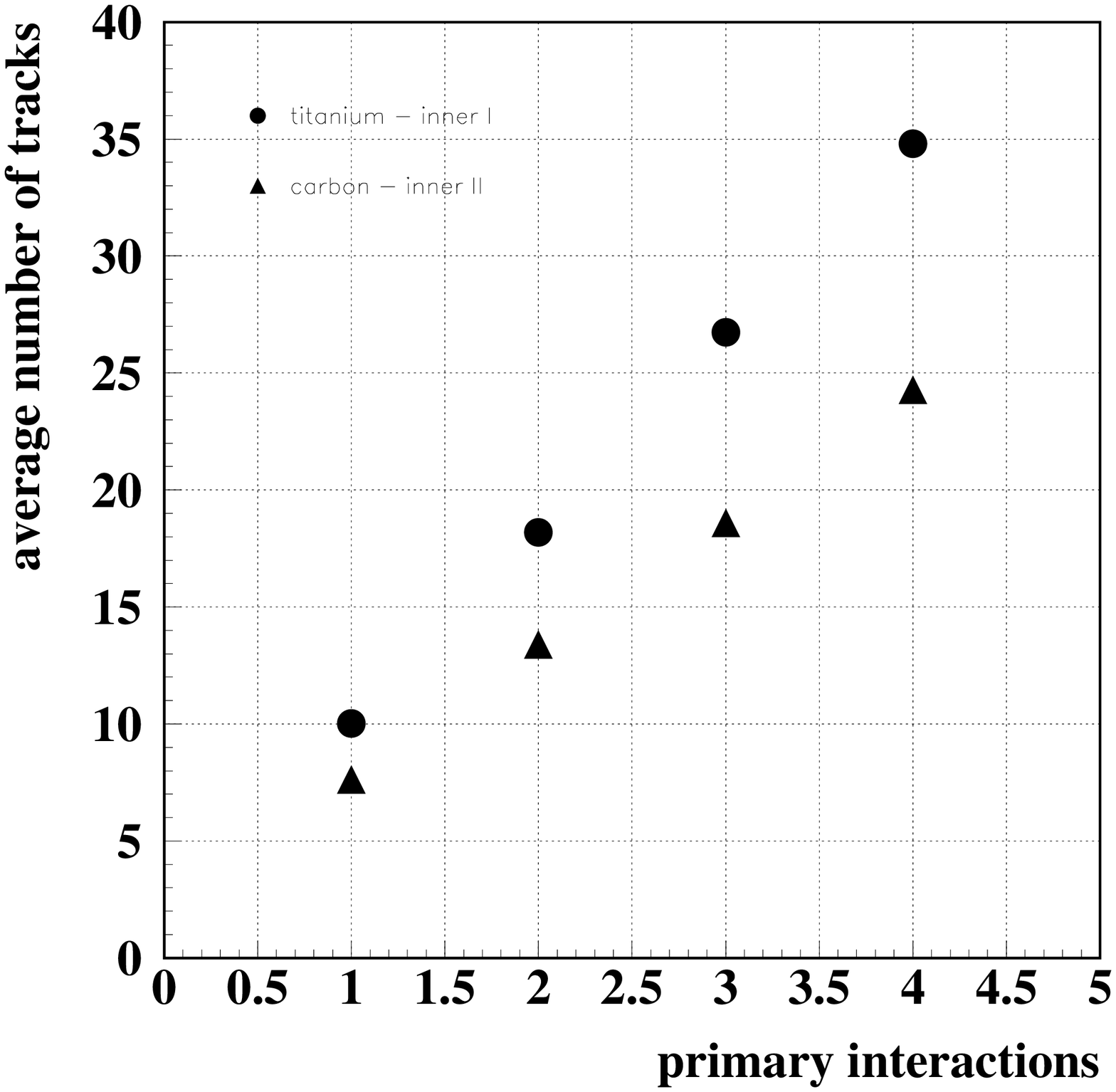, width=6cm}
\epsfig{file=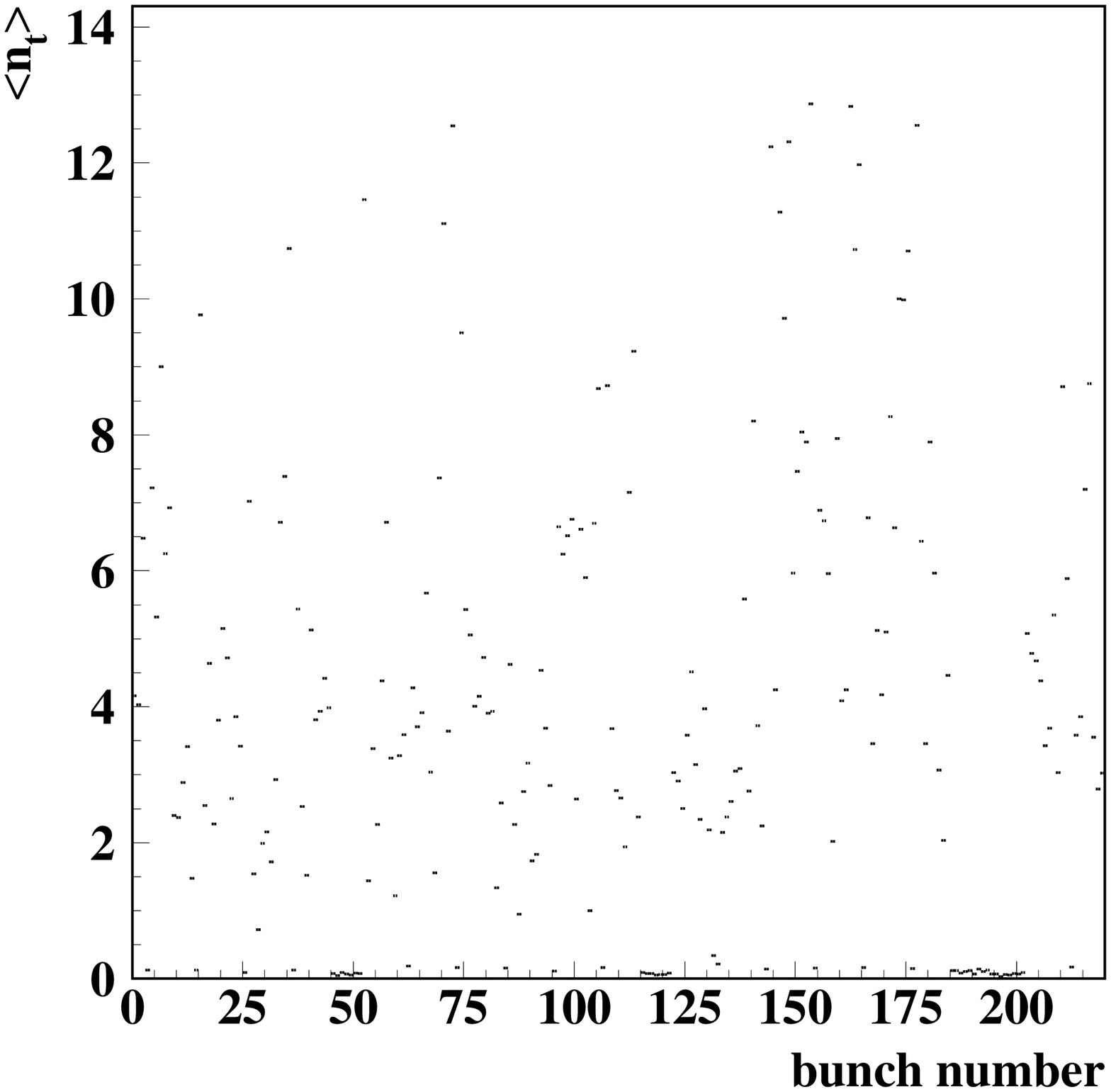, width=6cm}
\caption{Average number of reconstructed tracks as a function of the number of
superimposed primary interactions, given by a Monte Carlo simulation (left);
Average number of reconstructed tracks as a function of the bunch number for
a run taken with the carbon wire (right).}
\label{mcint}
\end{figure}

The plot on the left of \fig{mcint} shows a Monte Carlo simulation for the mean 
number of reconstructed tracks $\vev{n_t}_n$\/ which satisfy these criteria 
as a function of the number $n$\/ of superimposed interactions. One sees
that $\vev{n_t}_n$\/ indeed scales linearly with interaction rate up to 
4 superimposed interactions, which corresponds to a rate of about 40 MHz. 
Furthermore, heavier target materials yield higher track multiplicities.
The plot on the right of \fig{mcint} shows $\vev{n_t}_i$ as a function of the
bunch number $i$\/ for the same run considered in \fig{bxdist}. There are remarkable
variations in the track multiplicities between different bunches, which clearly
indicates distinct contributions to the total rate. In this plot, we can also 
identify the bunches which are nominally empty and contribute only 
marginally to the rate.

\subsection{Defining the systems}
%--------------------------------
In principle, any subdetector or combination of subdetectors in the experiment
can be chosen as a system for counting empty events. In order to minimize the 
dependence on Monte Carlo simulations, the requirement is a large acceptance 
for tagging non-empty events, reasonably low noise levels and good stability 
with time. We have chosen the most stable subdetectors in the data 
taking period of year 2000. System 1 consists of the vertex detector (VDS). 
An event is not empty in this system if:
\begin{itemize}
\item  there is at least 1 reconstructed track satisfying the track
selection criteria explained above.
\end{itemize}
System 2 is a combination of the ring imaging Cherenkov counter (RICH) and 
the electromagnetic calorimeter (ECAL). An event is considered to be not 
empty in this system if the following conditions are {\sl both} fulfilled:
\begin{itemize}
\item there are at least 30 reconstructed hits in the RICH.
\item the deposited energy in the inner part of the ECAL is above 5 GeV.
\end{itemize}

\begin{figure}[htb]
\centering
\epsfig{file=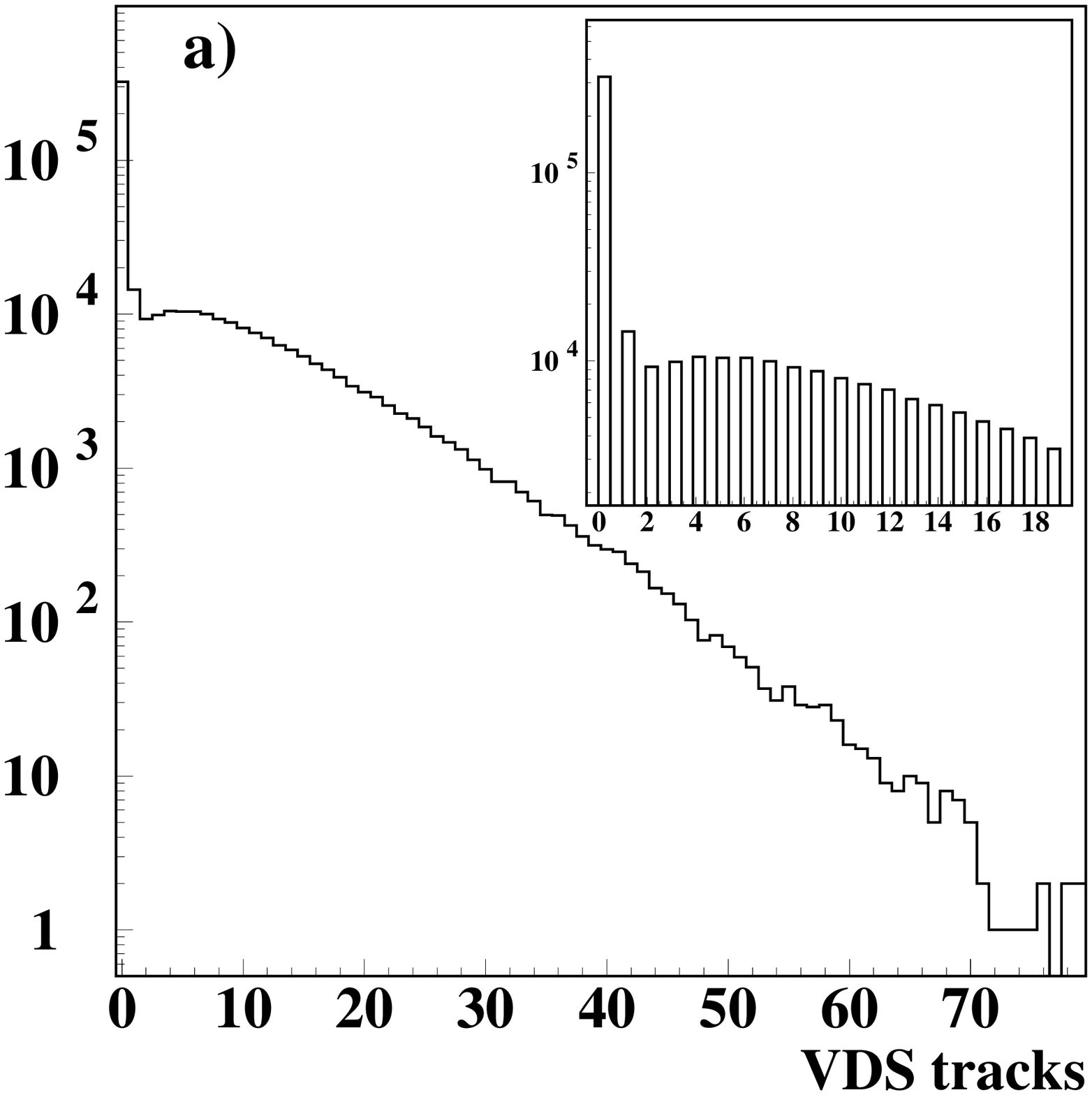, width=4.5cm}
\epsfig{file=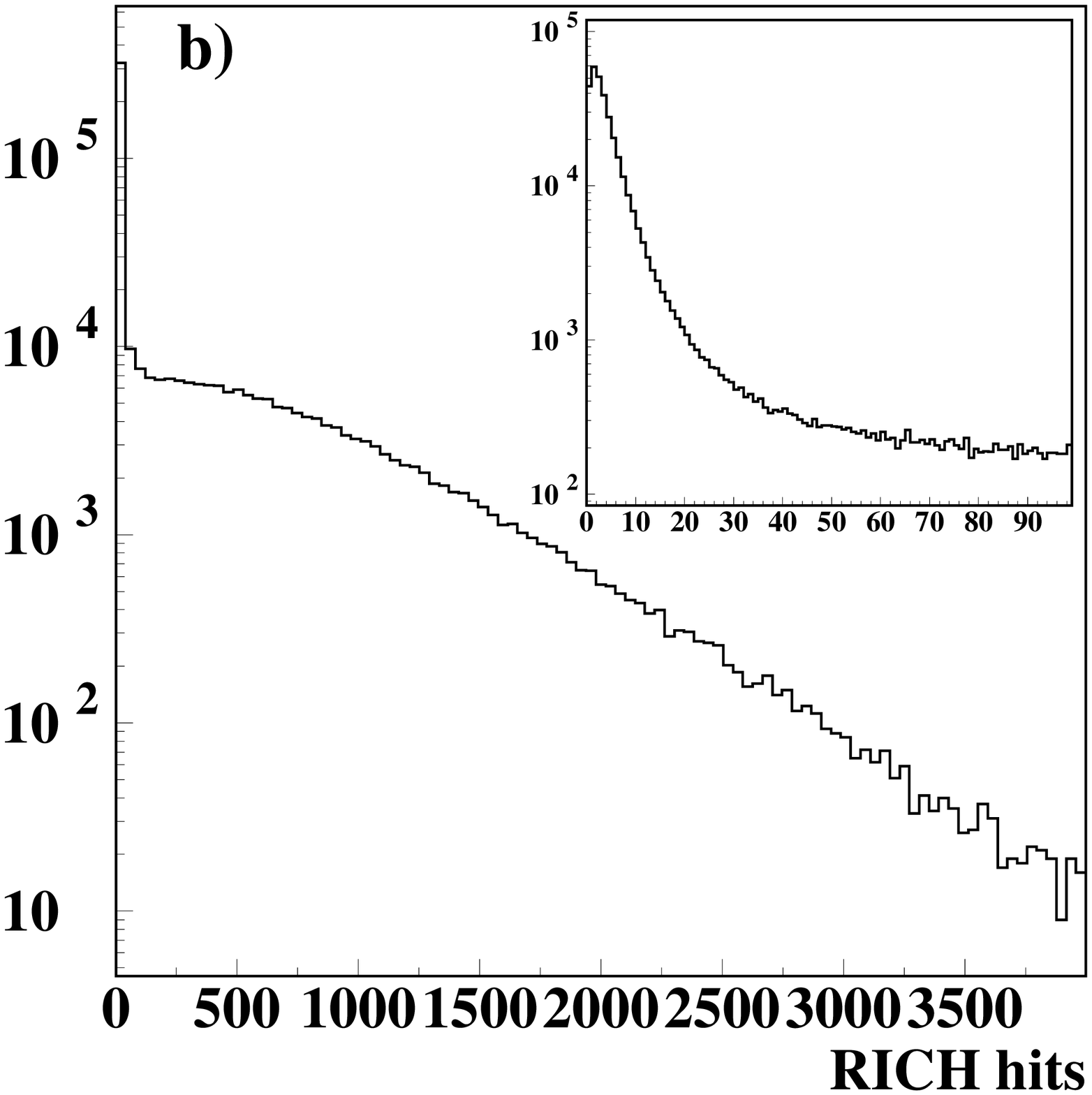, width=4.5cm}
\epsfig{file=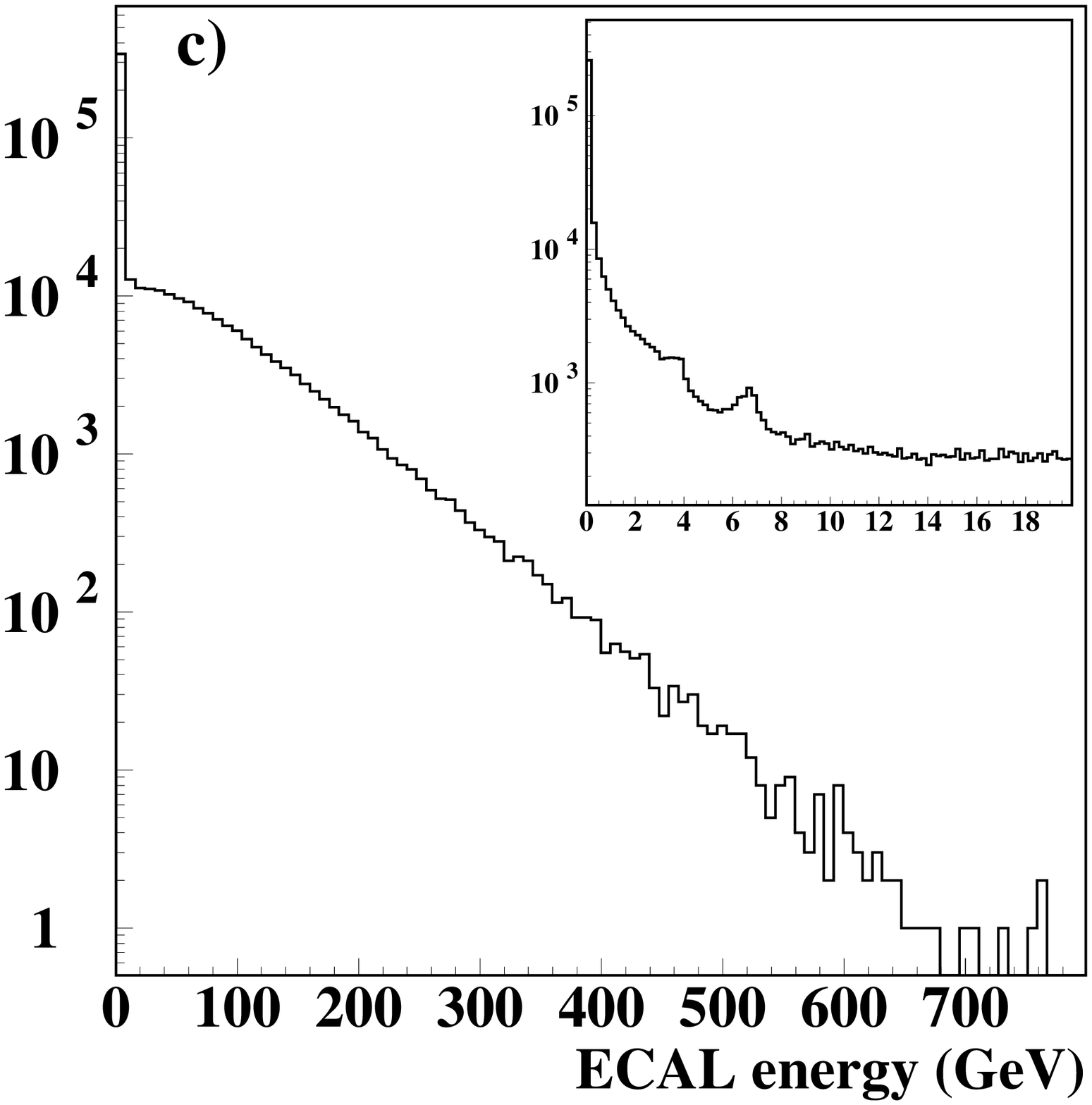, width=4.5cm}
\caption{Distributions of (a) number of tracks satisfying the
track selection criteria, (b) number of hits in the RICH,
and (c) energy deposition in inner ECAL, for a run taken with the
carbon wire. The inserts are a zoom to the first bins for each
distribution.}
\label{observ}
\end{figure}

In \fig{observ} we can see the distributions of number of tracks satisfying the
track selection criteria (a), number of hits in the Cherenkov detector
(b) and the total energy deposition in the inner part of the electromagnetic
calorimeter (c) for a run taken with a carbon target wire.

\begin{figure}[htb]
\centering
\epsfig{file=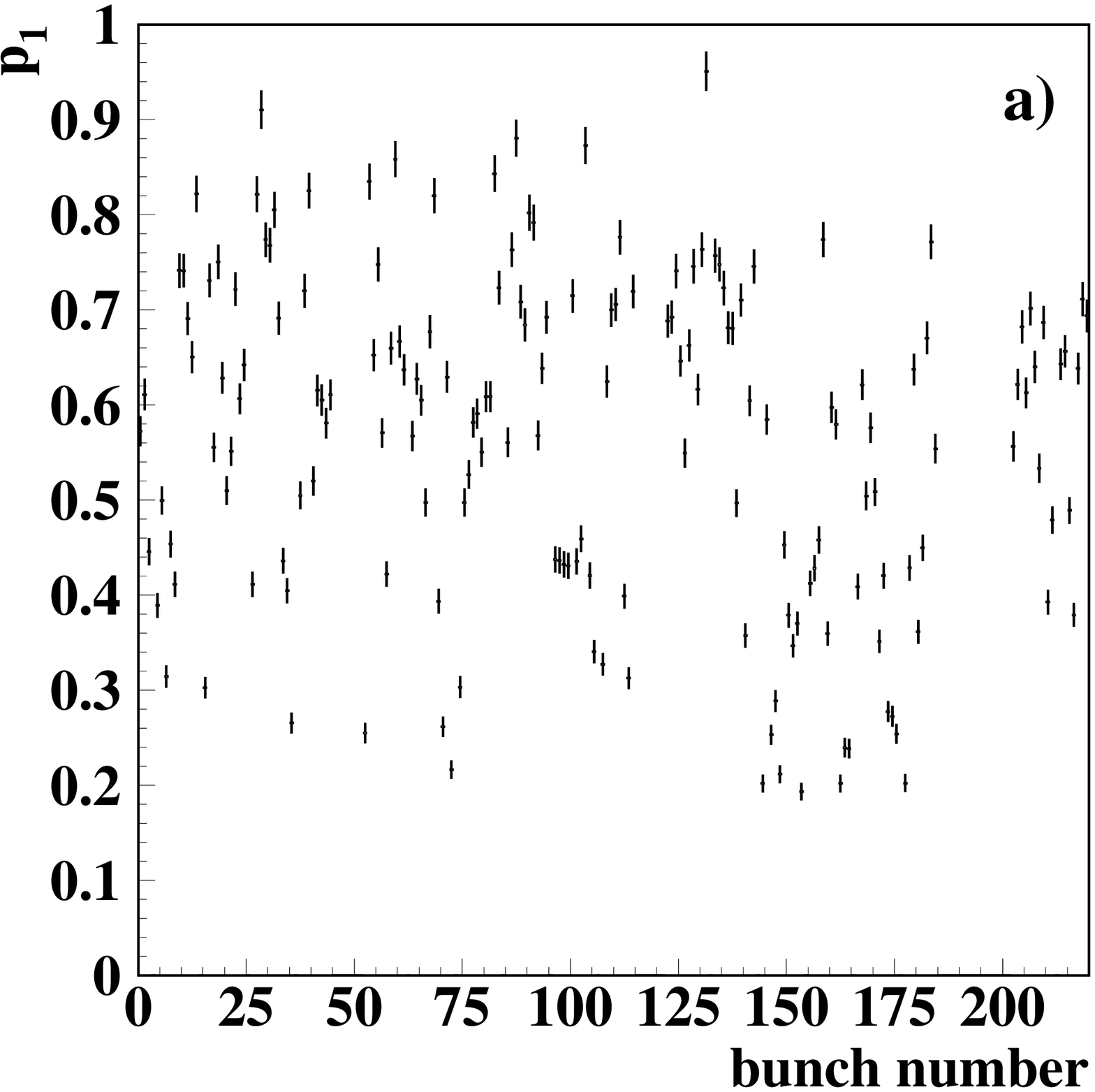, width=4.5cm}
\epsfig{file=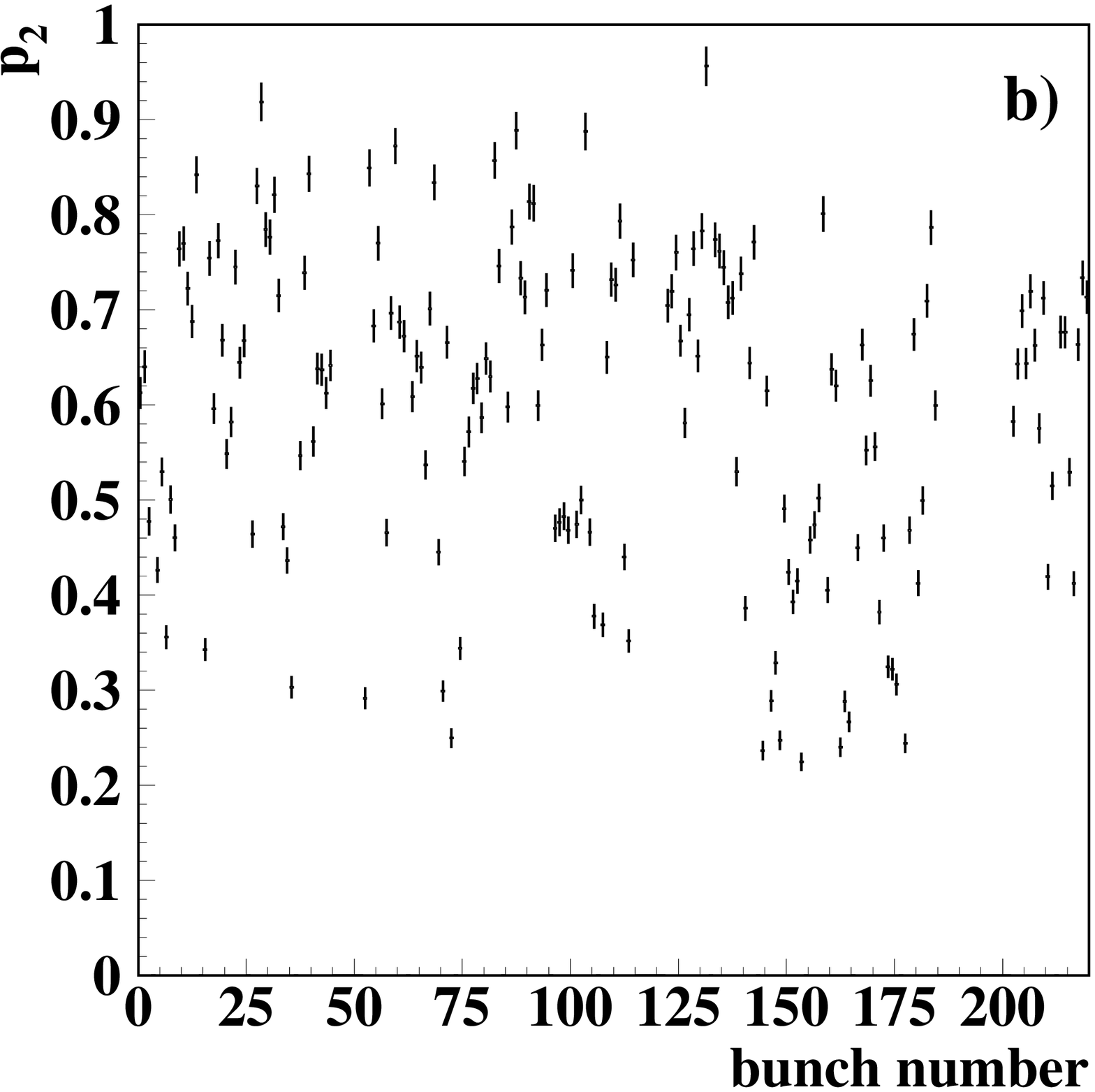, width=4.5cm}
\epsfig{file=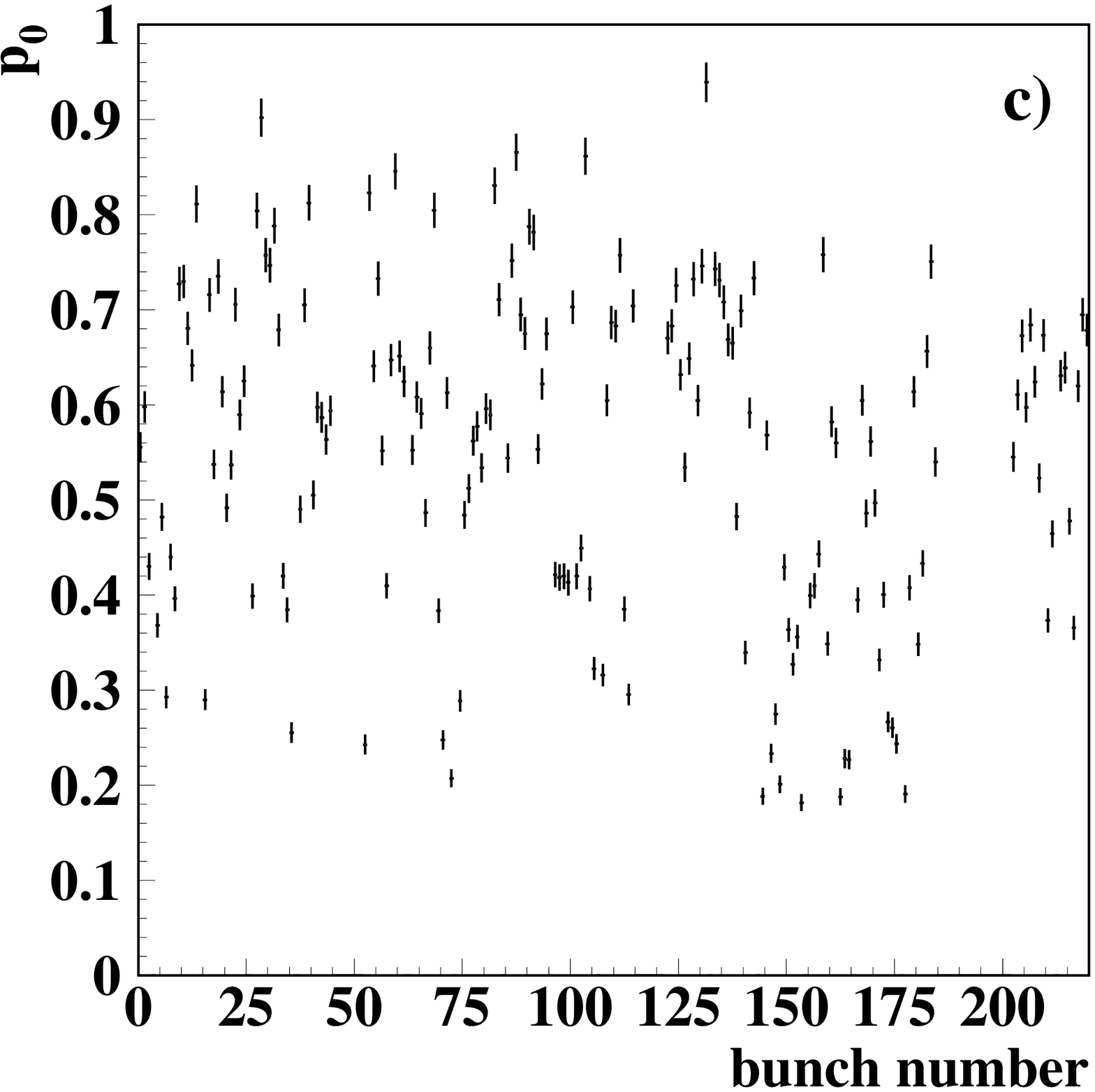, width=4.5cm}
\caption{Fraction of empty events as a function of bunch number
in (a) system 1; (b) system 2; and (c) both systems.}
\label{lnpk}
\end{figure}

In \fig{lnpk} we can find the probabilities $p_{ki}$, $k=\{0,1,2\}$, estimated 
as the fraction of empty events in system 1 (a), the fraction of empty 
events in system 2 (b) and the fraction of empty events in both systems 
(c), for the 180 nominally filled bunches. Again, remarkable variations 
are found between bunches, indicating different contributions to the 
total interaction rate.

\begin{figure}[htb]
\centering
\epsfig{file=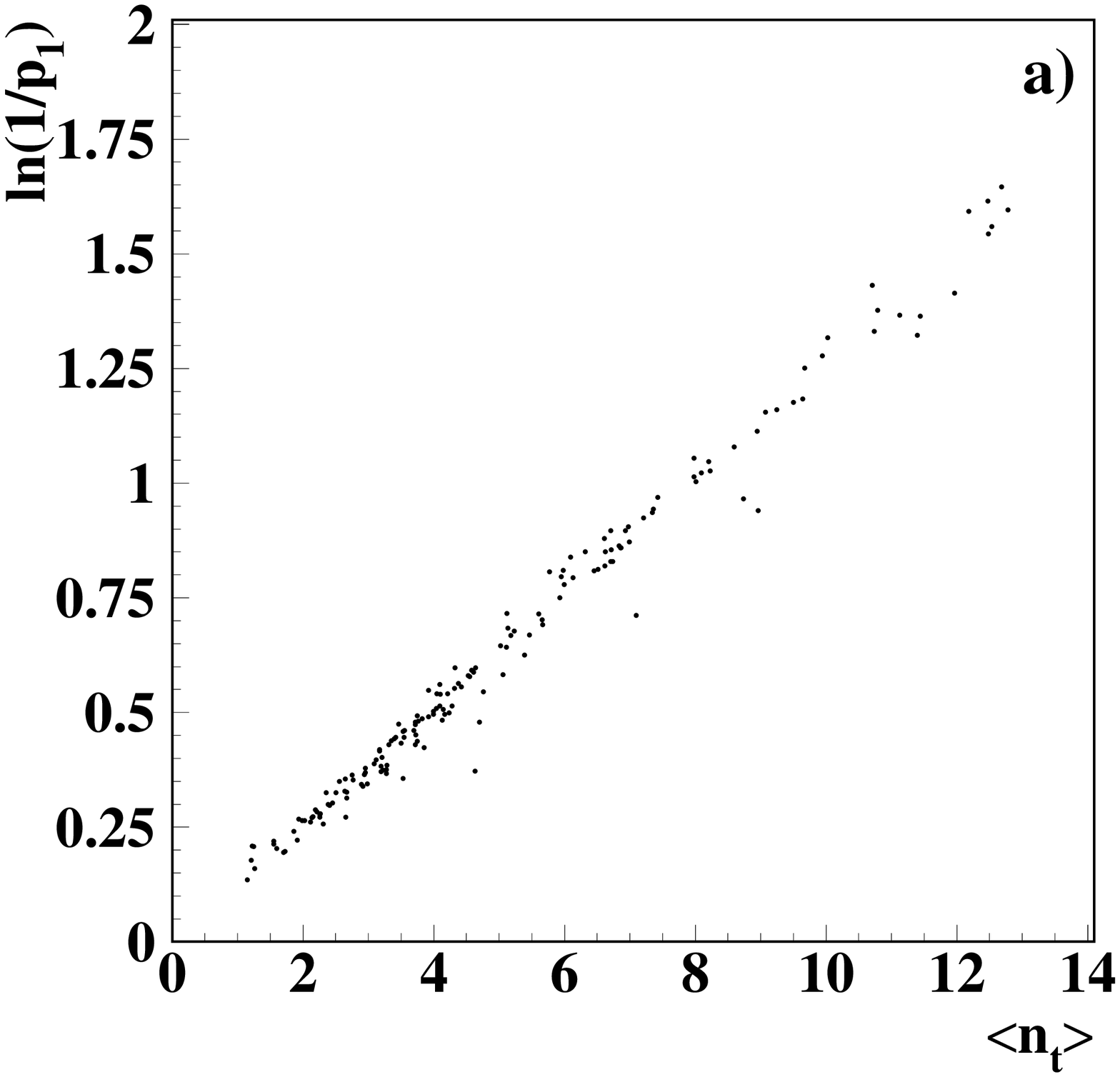, width=4.5cm}
\epsfig{file=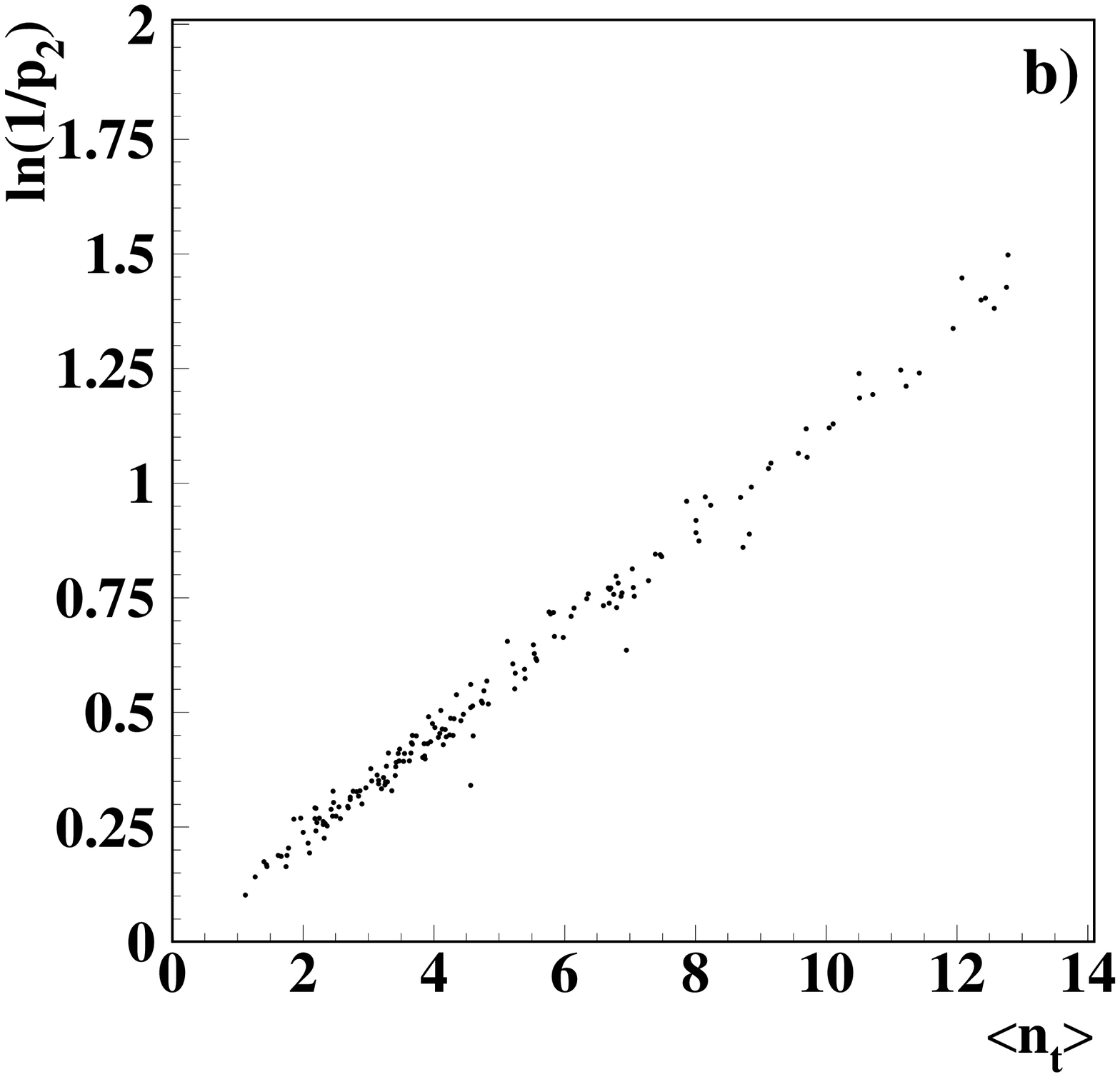, width=4.5cm}
\epsfig{file=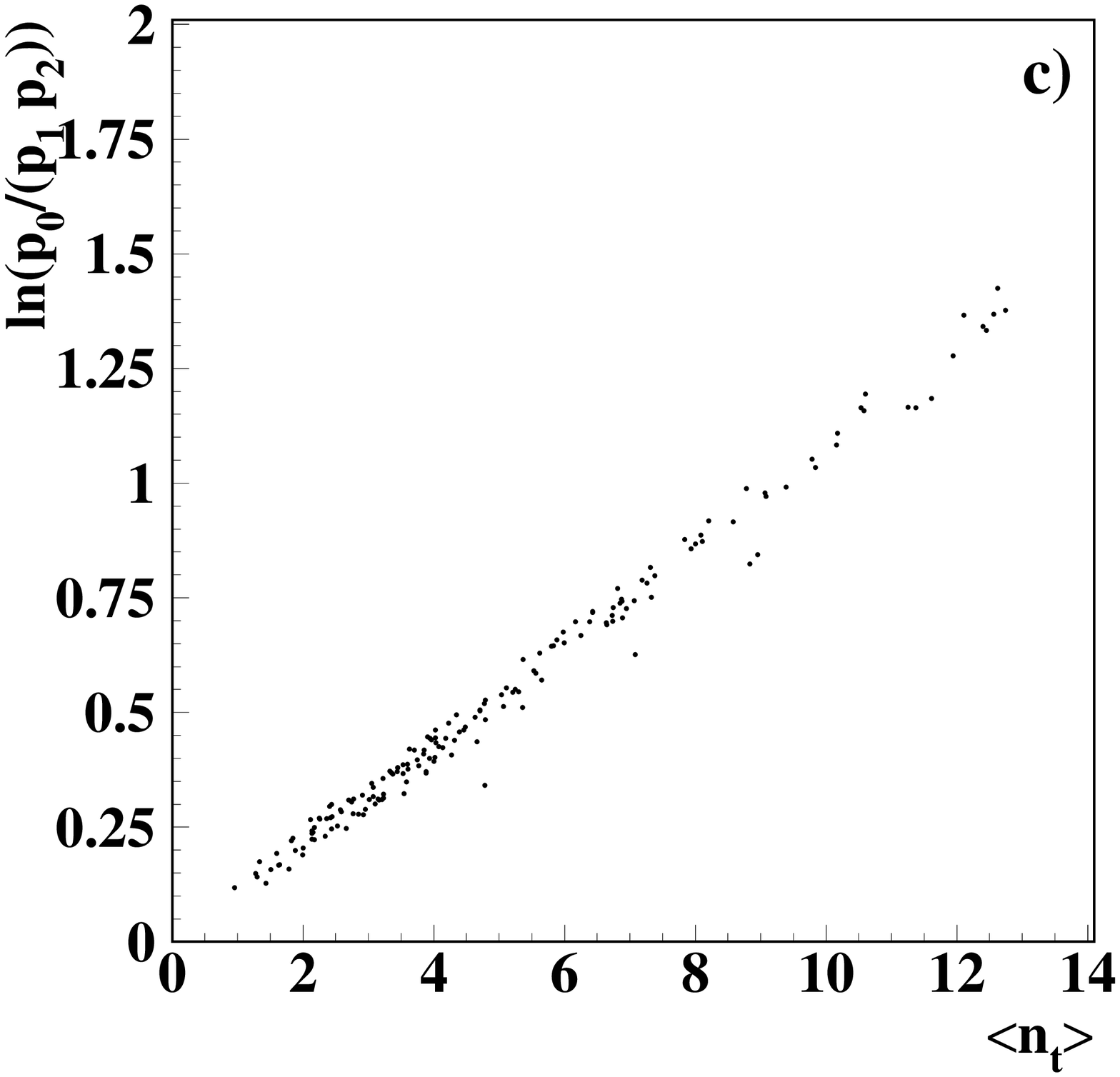, width=4.5cm}
\epsfig{file=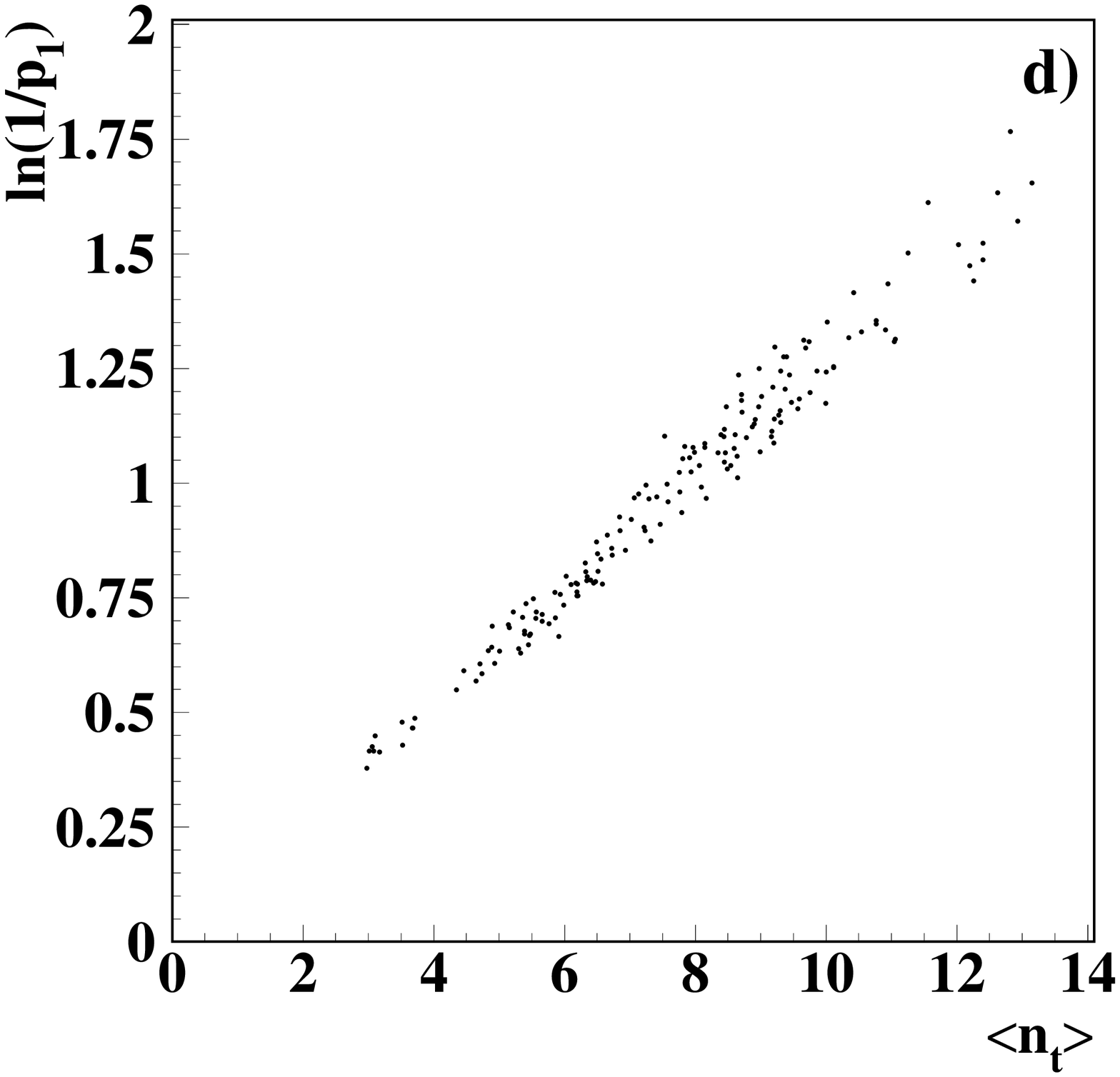, width=4.5cm}
\epsfig{file=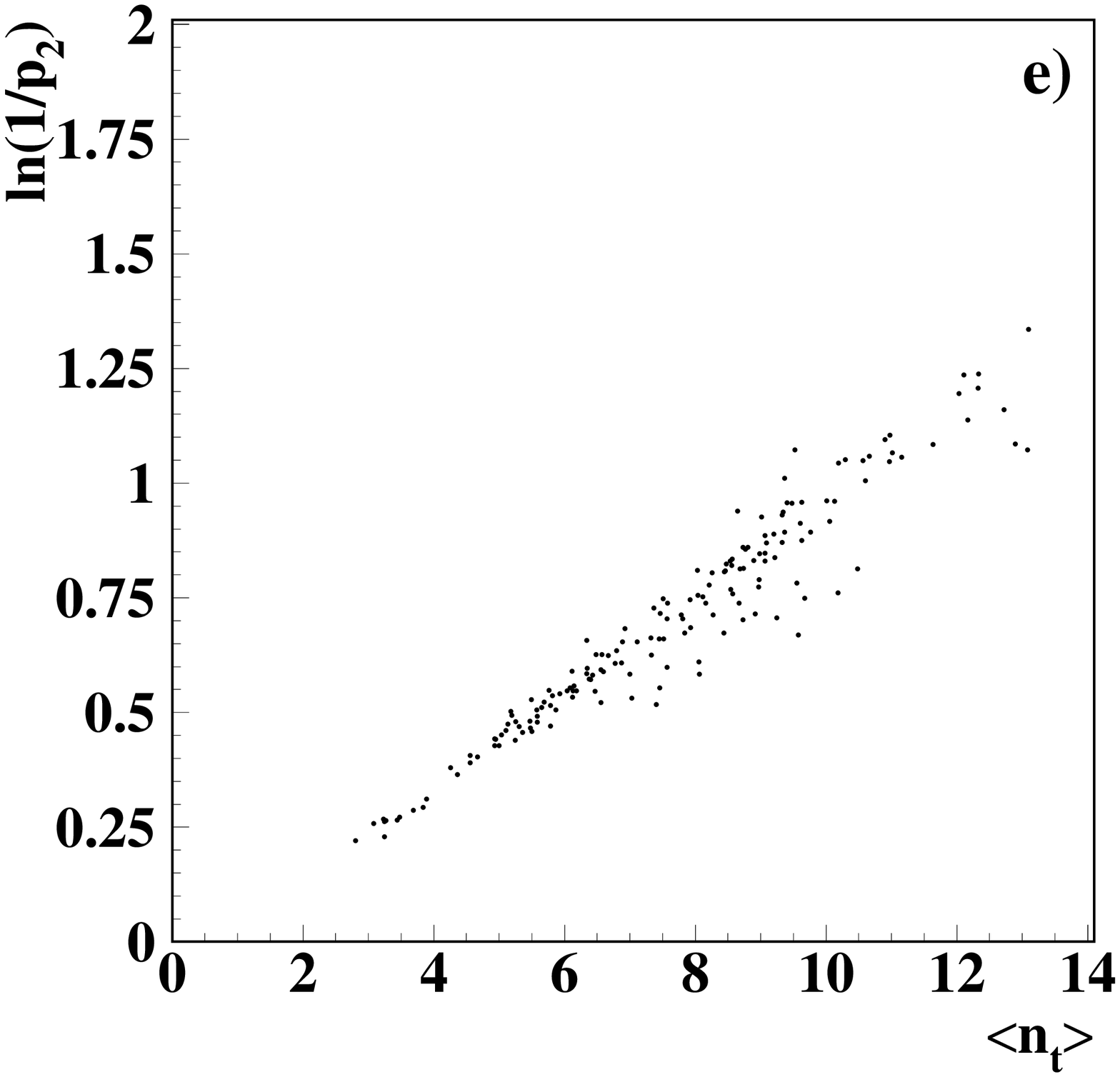, width=4.5cm}
\epsfig{file=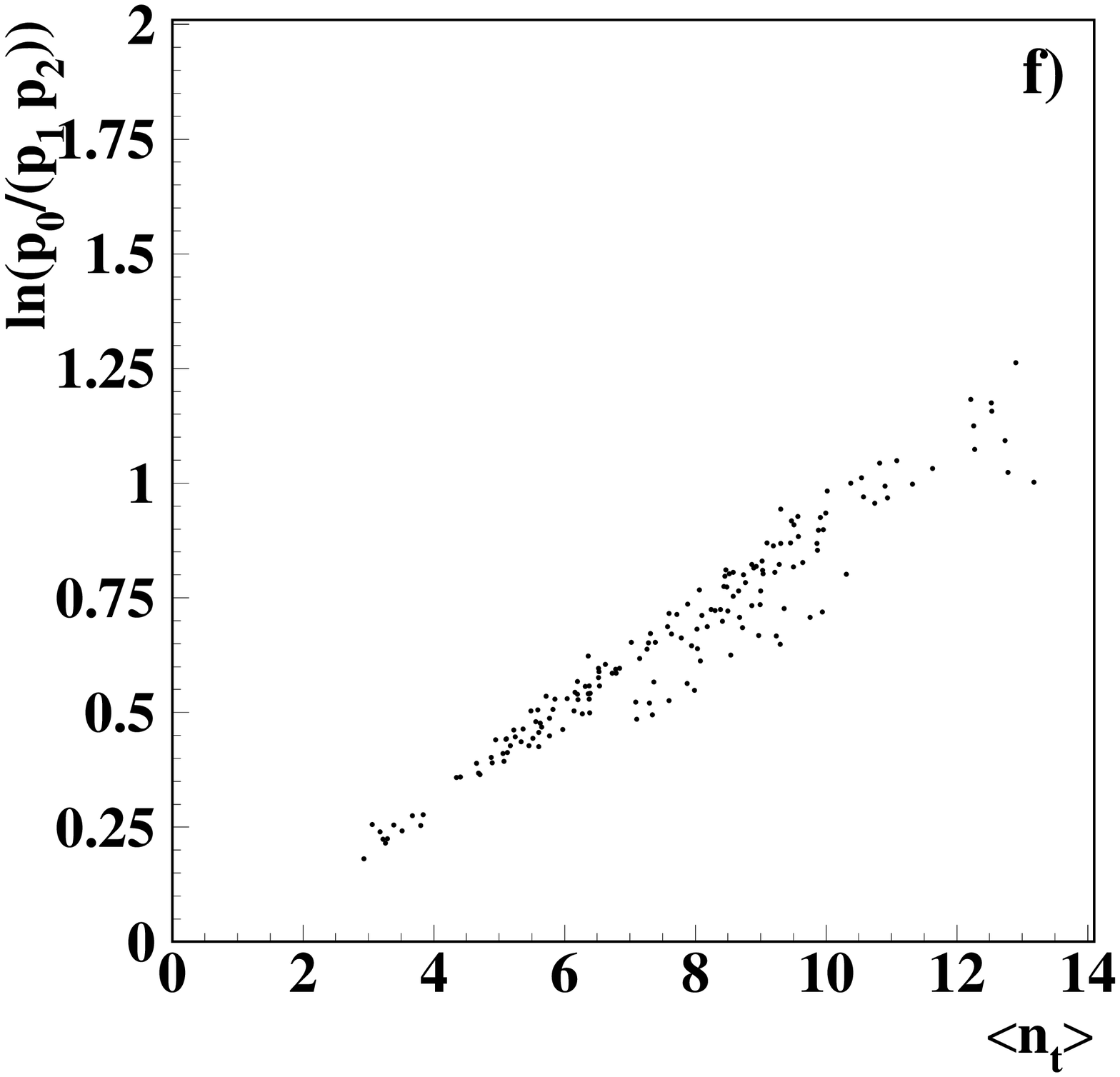, width=4.5cm}
\caption{Plots of $\ln(1/p_1)$, $\ln(1/p_2)$\/ and
$\ln(p_0/p_1p_2)$\/ as a function of $\vev{n_t}$. Each dot 
represents a bunch. The plots in the top row refer to a run taken 
with carbon target; the plots below refer to a run taken with 
aluminum target. The global parameters are obtained from a linear fit.}
\label{corrnt}
\end{figure}

\begin{figure}[htb]
\centering
\epsfig{file=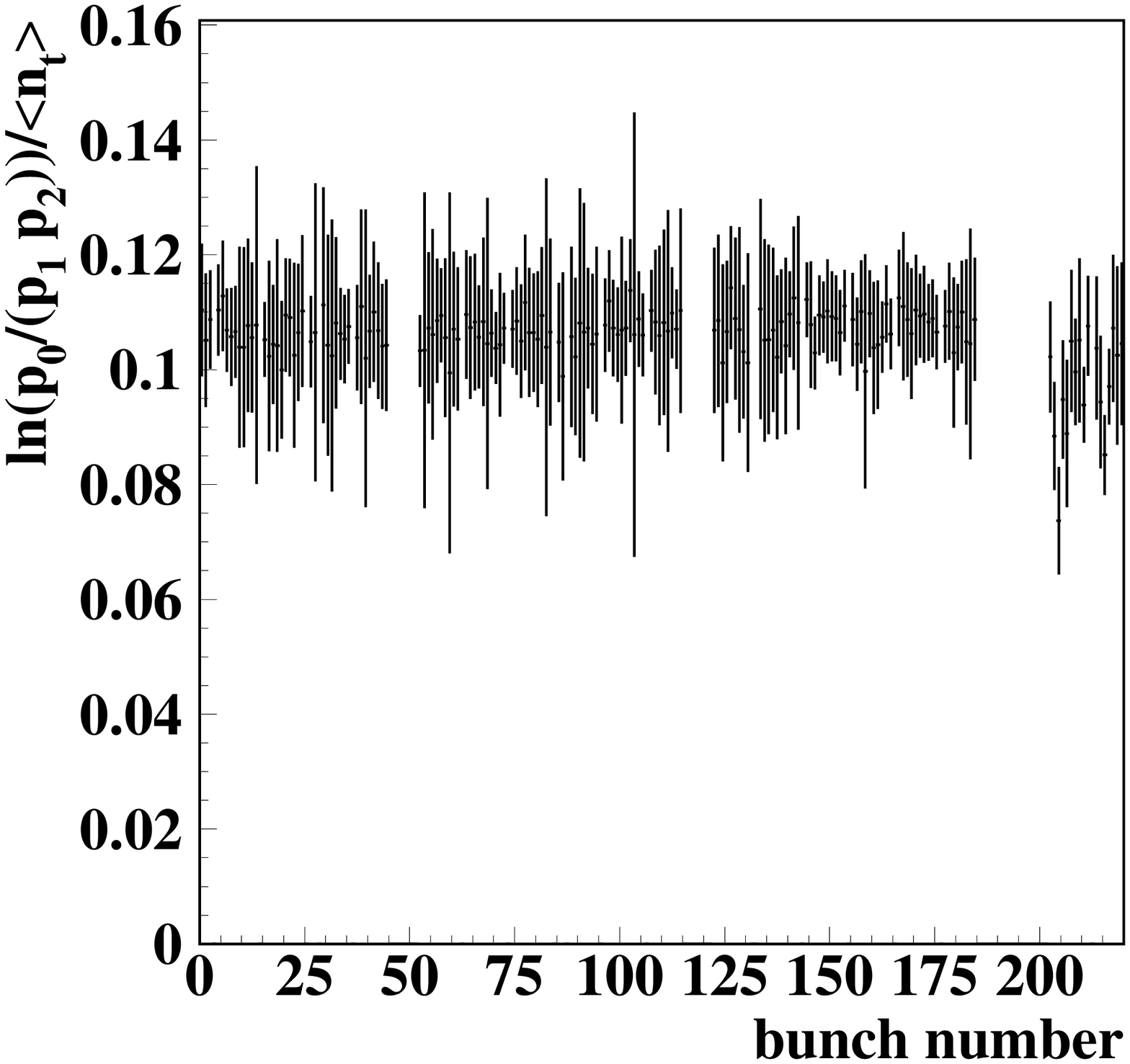, width=6cm}
\epsfig{file=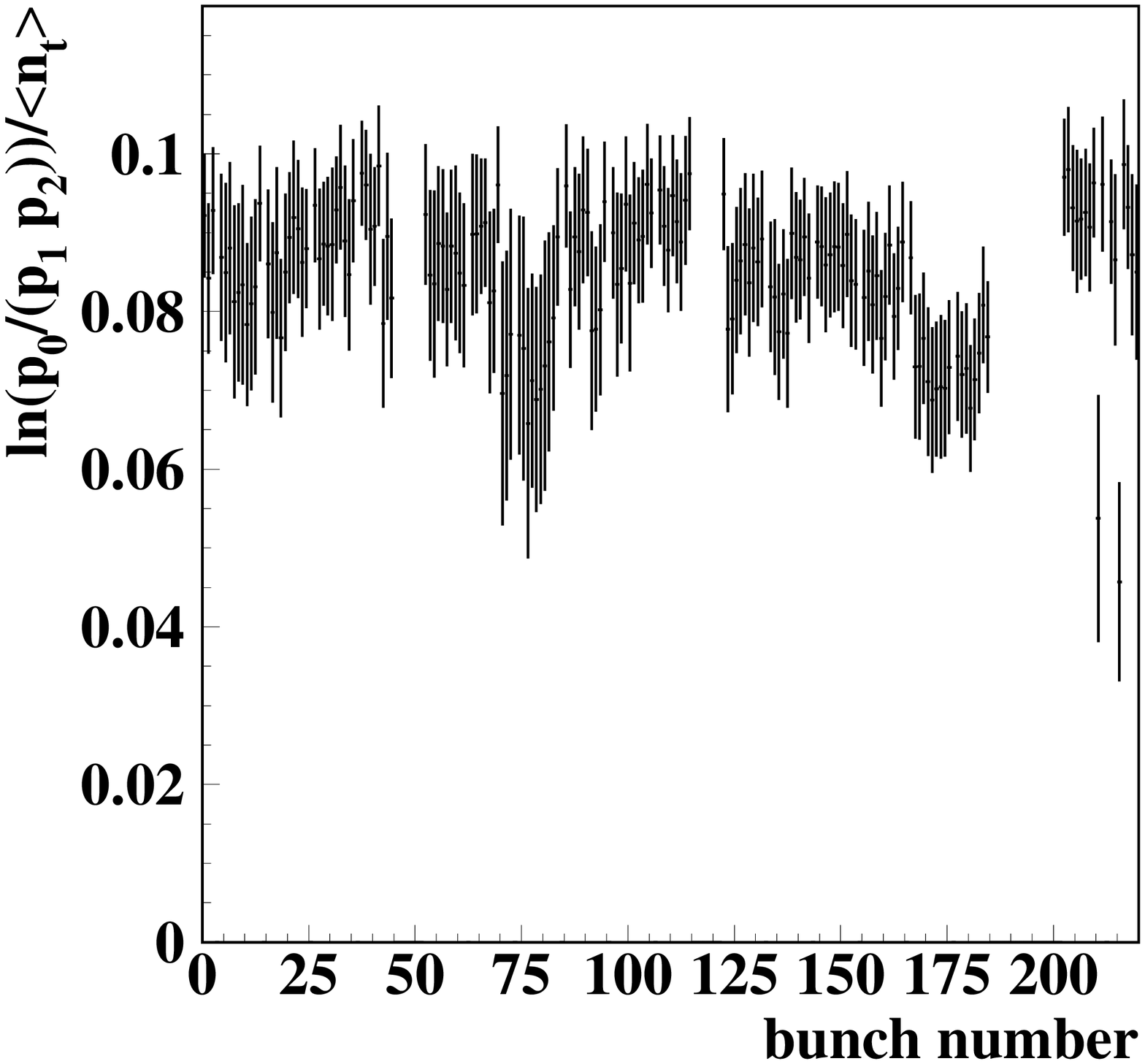, width=6cm}
\caption{Values of $\ln(p_0/p_1p_2)/\langle n_t \rangle$ as a function
of bunch number for a run taken with the carbon wire (left) and the
aluminum wire (right).}
\label{rfluct}
\end{figure}

\Fig{corrnt} shows $\ln(1/p_1)$, $\ln(1/p_2)$\/ and 
$\ln(p_0/p_1p_2)$\/ as a function of $\vev{n_t}$\/ for two 
different runs. Each entry corresponds to one bunch. The top row is for a 
run taken with the carbon wire and very small rate fluctuations. The 
bottom row corresponds to a run taken with an aluminum wire and large 
rate fluctuations. The global parameters are obtained from an unweighted 
linear fit, performed after the bunches subject to rate fluctuations have 
been removed from the fit. These bunches are identified according to the 
constraint on the probabilities $p_{ki}$ given by \eq{discr}. In 
\fig{rfluct} one can see the ratio $\ln(p_0/p_1p_2)/\vev{n_t}$, which should 
be approximately constant for negligible rate fluctuations. For the run 
taken with the carbon target wire (left plot) this ratio is reasonably 
constant, indicating the absence of significant rate fluctuations. 
On the other hand, for the run taken with the aluminum target wire there 
are bunches which are subject to deviations from Poisson statistics which 
can be identified by having a lower than average value for
$\ln(p_0/p_1p_2)/\vev{n_t}$. Notice that these bunches can also be 
identified in \fig{corrnt}(f) below the main line (where entries concentrate), 
since for a given $\vev{n_t}$\/ they will have a lower 
$\ln(p_0/p_1p_2)$.

In Table 1 we give the average values over all runs of the efficiencies 
$a_{1,2}$, the noise-probabilities $q_{1,2}$\/ and the average number of 
tracks per interaction $\tau$, obtained from the global fits to the selected 
runs. It can be seen that the efficiencies for system 1 are typically larger 
than for system 2. On the other hand, the efficiencies are, within errors, 
quite similar for all target materials. However, we could expect them to 
be larger for heavier materials, which yield higher track multiplicities.
The probabilities $q_{1,2}$ are similar for runs acquired with carbon, 
titanium and tungsten targets, but larger for runs acquired with the 
aluminum target. This fact may be explained by the large fraction of 
coasting beam (unbunched protons uniformly distributed under the pulsed 
bunch structure) which plagues all runs taken with this wire \cite{Ehret}.
Furthermore, because the runs taken with the aluminum wire target tend to
show large rate instabilities it is natural to speculate if these are 
related to the presence of coasting beam.

\begin{table}[t!]
\centering
\begin{tabular}{l  c  c  c  c  c  r} \hline
     & $a_1$         & $a_2$         & $q_1$             & $q_2$             &
     $\tau$         \\ 
\hline
C    & $0.95\pm0.02$ & $0.86\pm0.02$ & $0.0189\pm0.0003$ & $0.0116\pm0.0001$ &  $7.69\pm0.26$ \\
Al   & $0.93\pm0.02$ & $0.83\pm0.02$ & $0.0516\pm0.0008$ & $0.0535\pm0.0011$ &  $8.27\pm0.24$ \\
Ti   & $0.97\pm0.02$ & $0.86\pm0.02$ & $0.0212\pm0.0003$ & $0.0142\pm0.0002$ &  $9.95\pm0.25$ \\
W    & $0.96\pm0.06$ & $0.87\pm0.05$ & $0.0177\pm0.0002$ & $0.0126\pm0.0001$ &
     $13.23\pm0.99$ \\ 
\hline
\end{tabular}
\caption{Bunch independent variables obtained from global fits to
nominally filled bunches.}
\end{table}

The mean number of tracks per interaction $\tau$\/ increases, as expected,
with the atomic weight of the target material. This dependence is usually 
parameterized by a power law of the atomic weight: $\tau \propto A^{\beta}$.
If we fit the values of $\tau$ as a function of the target atomic weight $A$,
we obtain $\beta = 0.20 \pm 0.02$, which is statistically compatible with the 
result $\beta = 0.18 \pm 0.02$\/ obtained in an independent study employing 
the HERA-B vertex detector \cite{Perschke}.

\begin{figure}[htb]
\centering
\epsfig{file=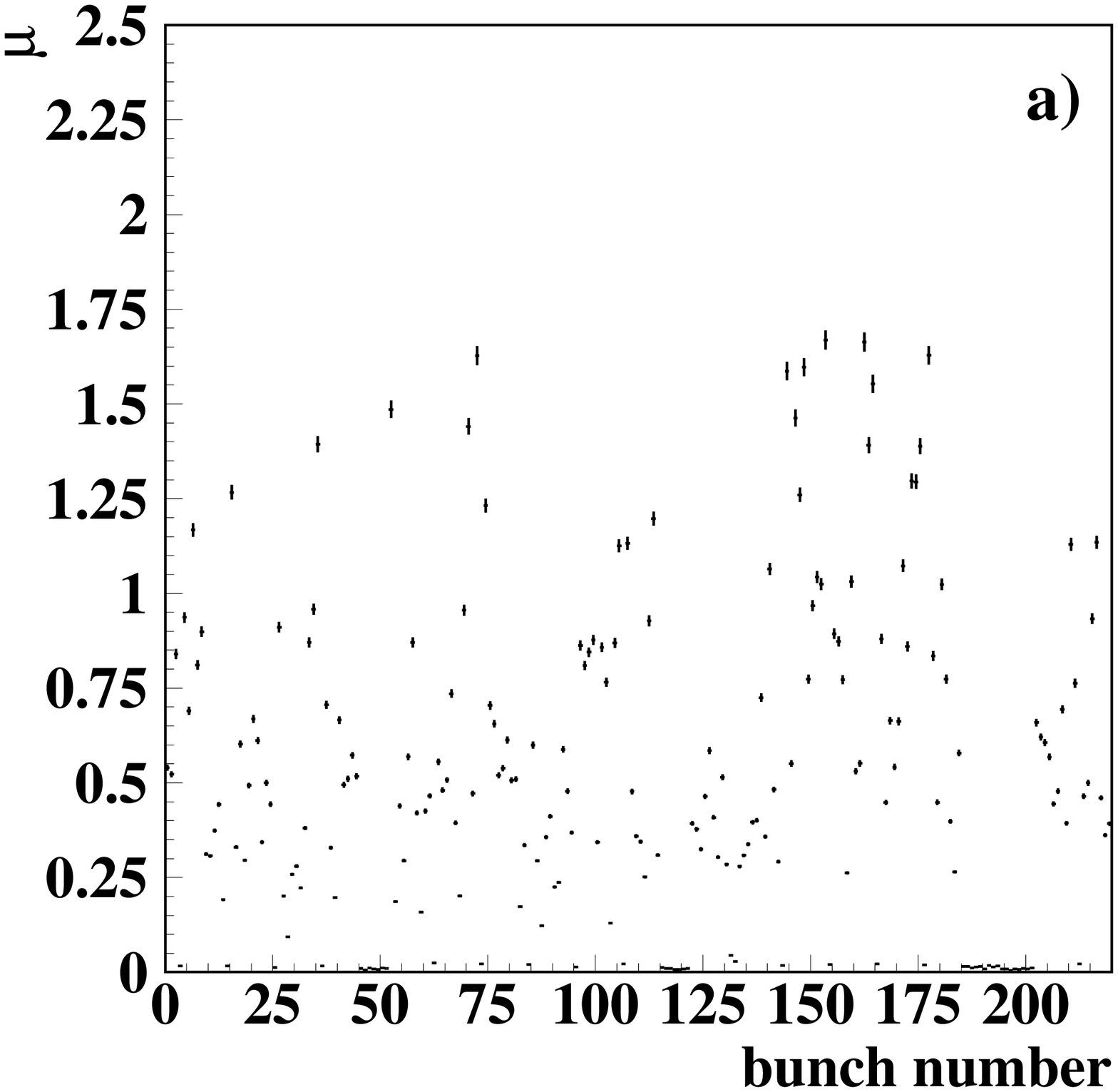, width=6cm}
\epsfig{file=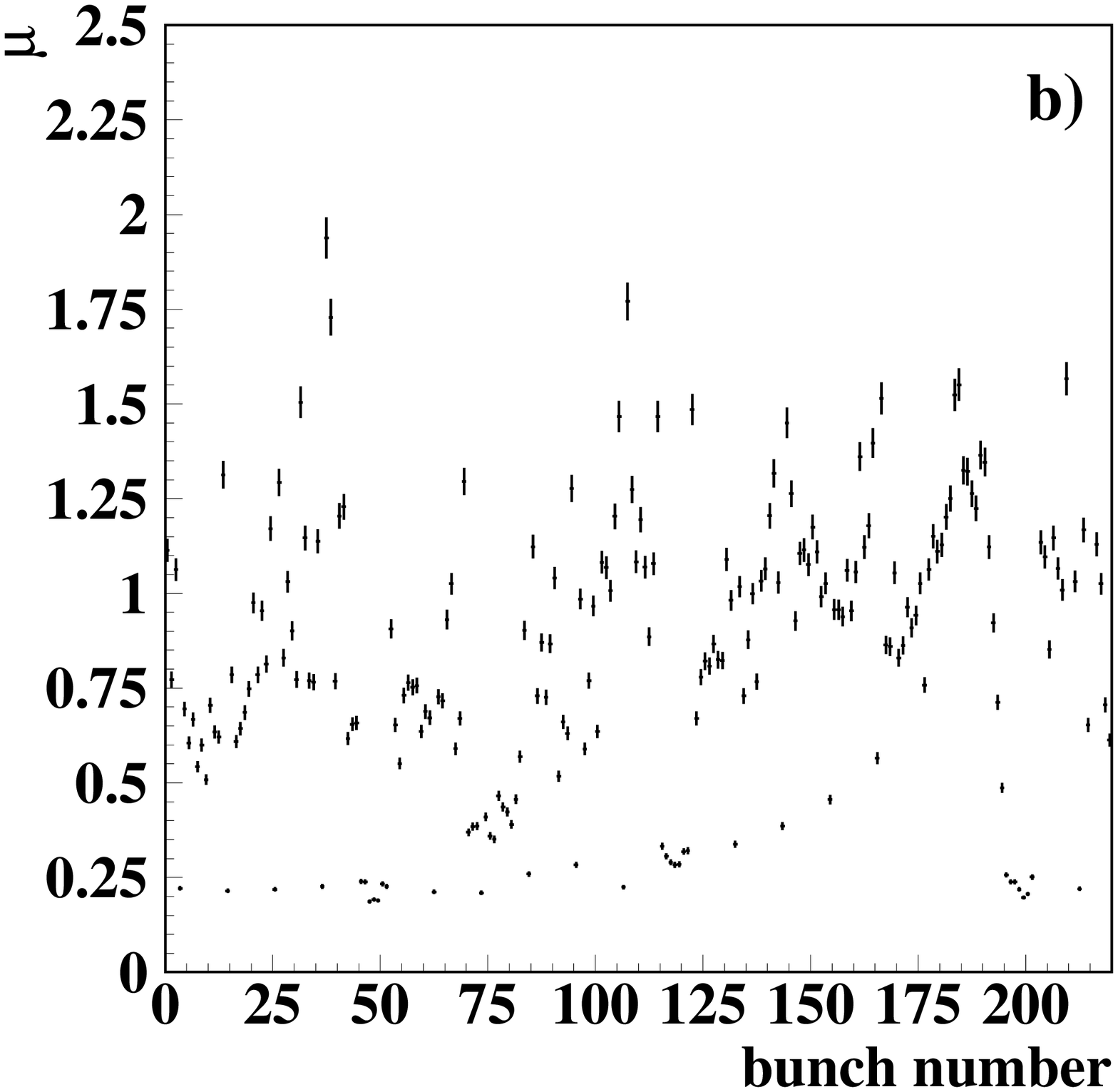, width=6cm}
\caption{Mean number of interactions as a function of bunch number for a run taken
(a) with a carbon target wire, and (b) with an aluminum target wire.
In plot (a) it can be seen the nominally empty bunches which contribute marginally
to the interaction rate. In plot (b) these bunches contribute significantly
to the rate, which is a clear indication of the high levels of coasting
beam affecting this run.}
\label{avint}
\end{figure}

Once $\tau$\/ is known, the average number of interactions per bunch crossing 
$\mu_i$\/ can be calculated according to \eq{taudef}. \Fig{avint} shows the 
values of $\mu_i$\/ for all bunches, for a run taken with the carbon target 
wire (a) and with the aluminum target wire (b). First, it is noteworthy 
that bunches contribute quite differently to the rate. In the run taken with
aluminum target we 
can see a large contribution of nominally empty bunches to the total rate. 
This behaviour can be observed in other runs taken with aluminum wire and, 
again, this can be explained by the large fraction of coasting beam which 
is present in all runs taken with this wire.

The total number of interactions $N_{int}$\/ in a run is given by
$N_{int} = \sum_{i=1}^{220}N_i\mu_i$, where $N_i$\/ is the total number of 
events due to bunch $i$. From $N_{int}$ the luminosity is obtained using 
\eq{lumidef} and the inelastic cross sections published in Ref. \cite{Carvalho}.

\subsection{Systematic uncertainties}
%------------------------------------
Because the final states of proton-nucleus interactions sample a 
large phase-space, certain event topologies may be outside the acceptance 
of both subdetectors, leading to systematic uncertainties in the measured 
luminosity. Events which are not seen by both systems do not contribute to
any inefficiency as inferred by the TSSM, and thus lead to an overestimate
of the true acceptance. The systematic uncertainties of the statistical method 
were studied with a toy Monte Carlo based on the interaction model MINT
\cite{MINT} and a coarse simulation of the HERA-B detector based on angular 
acceptance cuts, some rough estimates for the track finding efficiencies plus
some assumption about noise and smearing in the RICH and ECAL. The impact on 
the measured luminosity of diffractive contributions, rate fluctuations by 
$\pm 20\%$, target materials covering the range from Carbon to Tungsten
and nominal interaction rates varying by a factor $\pm e$\/ were considered. 
It is found that for a detector such as HERA-B, there is a small bias on 
the luminosity estimate from the TSSM. Assuming that the reference cross
section is the total inelastic cross section, the luminosity estimate is
between 3\% and 6\% too small. Using instead only the non-diffractive 
inelastic cross section as a reference, the results are between 1\% and 6\% 
too high. Taking conservatively the larger of the two ranges and correcting 
for the average bias, we conclude that the intrinsic systematic error of 
the TSSM is around 3\%. Note that this figure does not include systematic 
uncertainties due to imperfect knowledge of the contributing cross sections.

%%%%%%%%%%%%%%%%%%%%%%%%%%%%%%%%%%%%%%%%%%%%%%%%%%%%%%%%%%%%%%%%%%%%%%%%%%
%%                             CONCLUSIONS                              %%
%%%%%%%%%%%%%%%%%%%%%%%%%%%%%%%%%%%%%%%%%%%%%%%%%%%%%%%%%%%%%%%%%%%%%%%%%%

\section{Conclusions}
%--------------------
A statistical method to measure the integrated luminosity of high energy
interactions at collider experiments was presented. The method starts from 
the assumption that the number of interactions in a random triggered event 
follows Poisson statistics. Then, two large acceptance subdetectors of 
the experiment are considered. Counting the fraction of empty events in 
either of the two subdetectors and simultaneously in both, as function 
of the bunch crossing numbers, allows to infer the acceptance of the two 
subdetectors, noise contributions and total number of interactions from 
the data alone, thereby reducing the dependence of the analysis on Monte 
Carlo simulations. Introducing also information from an inclusive quantity, 
the method was implemented such that a bias due to rate fluctuations, which 
tend to spoil the assumption of Poisson statistics for the interaction 
multiplicity of a given bunch, can be avoided. This method was applied to 
random triggered minimum bias data collected in the commissioning  period of 
the HERA-B experiment in spring 2000. Without correcting the luminosity 
estimates for the bias caused by those parts of the cross section which 
are not seen by either of the two sub-systems considered, the TSSM would
have an intrinsic systematic error of 6\%. For more hermetic detectors 
and at higher energies even smaller uncertainties can be expected. 
Correcting for the bias, the intrinsic systematic error of the method
drops to 3\%.

\section*{Acknowledgment}
%-------------------------
We would like to thank our colleagues from the HERA-B 
collaboration for many useful discussions and their 
support in using HERA-B data to illustrate the method.
This work was supported by the Max-Planck Society and 
Funda\c c\~ao para a Ci\^encia e Tecnologia. One of us 
(JB) was covered by grant BD/16272/98.

% The Appendices part is started with the command \appendix;
% appendix sections are then done as normal sections
% \appendix

% \section{}
% \label{}


\begin{thebibliography}{00}

\bibitem{Hartouni} E. Hartouni \etal, HERA-B Technical Design Report, DESY-PRC 95/01.

\bibitem{marco} M. Bruschi, HERA-B 05-011, Physics 05-008, to be published.

\bibitem{Target} K. Ehret, Nucl. Instr. and Meth. A 446 (2000) 190.

\bibitem{VDS} C. Bauer \etal, Nucl. Instr. and Meth. A 453 (2000) 103.

\bibitem{ITR} T. Zeuner, Nucl. Instr. and Meth. A 446 (2000) 324.;
Y. Bagaturia \etal, Nucl. Instr. and Meth. A 490 (2002) 223.

\bibitem{OTR} M. Capeans, Nucl. Instr. and Meth. A 446 (2000) 317.

\bibitem{RICH} J. Pyrlik, Nucl. Instr. and Meth. A 446 (2000) 299.;
I. Ari\~no \etal,  Nucl. Instr. and Meth. A 453 (2000) 289.

\bibitem{ECAL} G. Avoni. \etal, Proc. of the IX Conference on Calorimetry in
Particle Physics, Annecy, France, October 9-14, 2000, Calorimetry in High Energy
Physics, (2001) 777.;
A. Zoccoli, Nucl. Instr. and Meth. A 446 (2000) 246.

\bibitem{MUON} M. Buchler \etal, IEEE Trans. Nucl. Sci. 46 (1999) 126.;
A. Arefiev \etal, IEEE Trans. Nucl. Sci. 48 (2001) 1059.

\bibitem{murthy} P.V.R. Murthy \etal, Nucl. Phys. B 92 (1975) 269.

\bibitem{bellettini} G. Bellettini \etal, Nucl. Phys. 79 (1966) 609.

\bibitem{carroll} A.S. Carroll \etal, Phys. Lett. B 80 (1979) 319.

\bibitem{fumuro} F. Fumuro \etal, Nucl. Phys. B 152 (1979) 376.

\bibitem{denisov} S.P. Denisov \etal, Nucl. Phys. B 61 (1973) 62.

\bibitem{roberts} T.J. Roberts \etal, Nucl. Phys. B 159 (1979) 56.

\bibitem{Carvalho} J. Carvalho, Nucl. Phys. A 725 (2003) 269.

\bibitem{FRITIOF} H. Pi, Comp. Phys. Comm. 71 (1992) 173.

\bibitem{Ehret} K. Ehret \etal, Nucl. Instr. and Meth. A 456 (2001) 206.

\bibitem{Perschke} T. Perschke, PhD Thesis, University of M\"unchen (2000).

\bibitem{MINT} M. Schmelling, Proc. of 40th Rencontres de Moriond on QCD and High Energy Hadronic
Interactions, La Thuile, Aosta Valley, Italy, 12-19 Mar 2005, hep-ph/0506028.


% \bibitem{label}
% Text of bibliographic item

% notes:
% \bibitem{label} \note

% subbibitems:
% \begin{subbibitems}{label}
% \bibitem{label1}
% \bibitem{label2}
% If there is a note, it should come last:
% \bibitem{label3} \note
% \end{subbibitems}


\end{thebibliography}
\end{document}